\documentclass[prd,aps,showpacs,,nofootinbib,superscriptaddress,preprintnumbers,epsf,psf]{revtex4}
\usepackage[english]{babel}
\usepackage{pstricks}
\usepackage[dvips]{graphicx}
\usepackage{epsf}
\usepackage{amsmath} 
\usepackage{amssymb}
\usepackage{psfrag}

\def\lsim{\raise0.3ex\hbox{$<$\kern-0.75em\raise-1.1ex\hbox{$\sim$}}}
\def\gsim{\raise0.3ex\hbox{$>$\kern-0.75em\raise-1.1ex\hbox{$\sim$}}}

\def\bei{\begin{itemize}}
\def\ei{\end{itemize}}
\def\beqa{\begin{eqnarray}}
\def\eqa{\end{eqnarray}}
\def\bea{\begin{eqnarray}}
\def\eea{\end{eqnarray}}
\def\beas{\begin{eqnarray*}}
\def\eeas{\end{eqnarray*}}
\def\beqas{\begin{eqnarray*}}
\def\eqas{\end{eqnarray*}}

\def\beq{\begin{equation}} 
\def\be{\begin{equation}}

\def\ee{\end{equation}}
\def\eq{\end{equation}}
\def\eeq{\end{equation}}

\def\beqd{\begin{displaymath}}
\def\eeqd{\end{displaymath}}
\def\eqd{\end{displaymath}}

\def\beeq{\begin{eqnarray}} \def\eeeq{\end{eqnarray}}

\def\bef{\begin{frame}}

\def\slashchar#1{\setbox0=\hbox{$#1$}
   \dimen0=\wd0
   \setbox1=\hbox{/} \dimen1=\wd1
   \ifdim\dimen0>\dimen1
      \rlap{\hbox to \dimen0{\hfil/\hfil}}
      #1
   \else
      \rlap{\hbox to \dimen1{\hfil$#1$\hfil}}
      /
   \fi}

\usepackage{graphicx}
\usepackage{dcolumn}
\usepackage{bm}
\usepackage{epsfig}

\newcommand{\dhd}{{\textstyle d}
\lower.03ex\hbox{\kern-0.38em$^{\scriptstyle-}$}\kern-0.05em{}}
\newcommand{\dbar}{{\textstyle \delta}
\lower.03ex\hbox{\kern-0.38em$^{\scriptstyle-}$}\kern-0.05em{}}
\newcommand{\half}{{1\over 2}}

\newcommand{\calu}{{\cal U}} 
\newcommand{\calv}{{\cal V}}

\newcommand{\hb}{\bar{h}} 

\begin{document}


\newcommand{\zbar}{\bar{z}}
\newcommand{\odd}{\mathbb{O}}
\newcommand{\pom}{\mathbb{P}}
 \makeatletter
 \def\preprint#1{ \def\@preprint{\noindent\hfill\hbox{#1}\vskip 10pt}}


\noindent\hfill\hbox{\begin{tabular}{l}
      CPHT-RR082.0910 \\
       LPT 10-73 \\
INT-PUB-10-047
 \end{tabular}}\vskip 10pt



\title{Uncovering the triple $\pom$omeron vertex from Wilson line formalism}

\author{G. A. Chirilli}
\affiliation{LPT, Universit{\'e} Paris-Sud, CNRS, 91405, Orsay, France
 {\em \&} \\
CPHT, {\'E}cole Polytechnique, CNRS, 91128 Palaiseau Cedex, France}

\author{L. Szymanowski}
\affiliation{Soltan Institute for Nuclear Studies, PL-00-681 Warsaw, Poland}

\author{S. Wallon}
\affiliation{LPT, Universit{\'e} Paris-Sud, CNRS, 91405, Orsay, France {\em \&} \\
UPMC Univ. Paris 06, facult\'e de physique, 4 place Jussieu, 75252 Paris Cedex 05, France}

\begin{abstract}
We compute the triple $\pom$omeron vertex from the Wilson line formalism, including both planar and non-planar contributions, 
and get perfect agreement with the result obtained in the Extended Generalized Logarithmic Approximation based on reggeon calculus.
\end{abstract}
%


\maketitle
\narrowtext


\noindent
\section{ Introduction}

It is well known that the Froissart bound \cite{Froissart:1961ux} for total cross-section, $\sigma_{tot} \leq \pi/m_{\pi}^2\ln^2 s$ \cite{Lukaszuk:1967zz}, is violated in perturbation theory within
the Leading Logarithmic Approximation (LLA).
The hard $\pom$omeron exchange obtained in the
LLA is described by the BFKL equation \cite{Fadin:1975cb,Kuraev:1976ge,Kuraev:1977fs,Balitsky:1978ic}. 
This violation persists also at the Next to Leading Logarithmic Approximation (NLLA).
Although strictly speaking valid only for hadronic observables, and not for external virtual states such as $\gamma^*,$ there is a common belief that
this bound should
be satisfied within any reasonable perturbatively resummed scheme.
This was the starting point of various lines of research, which led to various unitarization and saturation 
models in which the triple $\pom$omeron vertex is a key building block.

The {\it Generalized Leading Log Approximation}, which takes into account any {\it fixed} number $n$ of
$t$-channel exchanged reggeons, leads to the 
Bartels, Jaroszewicz, Kwiecinski, Praszalowicz (BJKP) equation \cite{Bartels:1978fc, Bartels:1980pe, Jaroszewicz:1980mq, Kwiecinski:1980wb}. The BJKP equation can be reformulated  as a 2-dimensional quantum mechanical problem
 with $n$ sites, each one corresponding to one of the (gluonic) $t-$channel reggeons (in the coordinate space), with time $\sim \ln s$.
The underlying hamiltonian is holomorphically separable  and invariant under global conformal transformations \cite{Lipatov:1990zb, Lipatov:1993qn}.
In the large $N_c$ limit, this 2-dimensional quantum mechanical model greatly simplifies, and turns out to be integrable
 \cite{Lipatov:1993qn, Lipatov:1993yb, Lipatov:1994xy, Faddeev:1994zg, Korchemsky:1994um}.
The singlet color bound states of an even number $n$ of $t-$channel reggeons have the quantum numbers of the $\pom$omeron (with $P\!=\!C=\!+1$), while for $n$ odd, such bound states contribute both to $\pom$omeron and Odderon (with $P\!=\!C=\!-1$) exchange.
For Odderon, the obtained trajectory satisfies $\alpha_\odd <1$ \cite{Janik:1998xj, Korchemsky:2001nx, Derkachov:2001yn, Derkachov:2002wz, DeVega:2001pu, deVega:2002im}. When summing with respect to $n$, it is is expected that the whole series, although divergent, could have a critical behavior with  an Odderon intercept $\alpha_\odd =1$.
However, these bound states       
  decouple from Born impact factors. They couple to photon impact factor
only through non trivial color states, at least of quadrupole type, which are therefore suppressed by $1/N_c^2$ powers. 
In contrast, 
 it is possible to exhibit a {\it critical} solution ($\alpha_\odd =1$)   which couples to Born impact
	       factors. These peculiar solutions
            can be obtained
              either from the perturbative Regge approach
 \cite{Bartels:1999yt}
             or from the dipole model, see Ref.~\cite{Kovchegov:2003dm}.

In comparison to the previous approach, the {\it Extended Generalized Leading Log Approximation} (EGLLA) \cite{Bartels:1991bh, Bartels:1992ym, Bartels:1993ih, Bartels:1994jj, Bartels:1999aw}, in which the number of reggeon in
       $t-$channel is not conserved, satisfies full unitarity (in all sub-channel) and leads to an
{\it effective 2-d field theory} realizing the Gribov  idea of Reggeon field theory \cite{Gribov:1968fc} in QCD (for a pedagogical review on this approach see Ref.~\cite{Bartels:1999aw}).
 In the framework of EGLLA, the simplest new building block (with singlet sub-channels) is the triple $\pom$omeron vertex 
\cite{Bartels:1993ih, Bartels:1992ym, Bartels:1994jj, Bartels:1995kf}. The conformal properties of this vertex allow one to relate it to the conformal blocks of an underlying (still unknown) conformal field theory, and using bootstrap properties, 
it was possible to evaluate this vertex \cite{Korchemsky:1997fy}.
The $\pom$omeron vertex contains two contributions: a planar one, and a non-planar one,
which is suppressed by a factor of $1/N_c^2$ with respect to the planar one:
\beq
\label{defV}
V^{1 \pom \to 2 \pom}=V^{1 \pom \to 2 \pom}_{\rm planar}+V^{1 \pom \to 2 \pom}_{\rm non-planar}\,.
\eq

The dipole model \cite{Nikolaev:1990ja, Nikolaev:1994vf}, \cite{Mueller:1993rr, Mueller:1994jq, Mueller:1994gb, Chen:1995pa}, equivalent to the BFKL equation at LLA \cite{Chen:1995pa, Navelet:1997tx}, is based on the description of the wave function of an onium state in terms of color dipole degrees of freedom in the 't Hooft limit and at large $s$. This wave function satisfies a non linear evolution equation. 
In this approach the vertex of $1\to 2$ dipoles \cite{Peschanski:1997yx, Bialas:1997ig, Bialas:1997xp}
is equivalent to the planar part of the triple $\pom$omeron vertex derived through the EGLLA approach. The non-planar part of the triple $\pom$omeron vertex cannot be derived from the dipole model since this approach relies on the large $N_c$ limit which suppresses all non-planar contributions. 

Since the time it became clear that at high-energy (Regge limit) non-linear effects dominates the dynamics of the 
scattering processes, non-linear evolution equations were derived. One of these equations is the Balitsky-Kovchegov (BK) 
equation, derived first by Balitsky \cite{Balitsky:1995ub, Balitsky:1998kc, Balitsky:1998ya, Balitsky:2001re}
in the Wilson line formalism, and then by Kovchegov \cite{Kovchegov:1999yj, Kovchegov:1999ua} in the dipole model.

The Wilson line formalism is an operator language. It is based on the concept of
factorization of the scattering amplitude in rapidity space and on the extension
to high-energy (Regge limit) of the Operator Product Expansion (OPE) technique, 
which was known before only at moderate energy (Bjorken limit) as an expansion in terms of local operators or in terms of light-ray operators. 
In Deep Inelastic Scattering (DIS) off a hadron at high-energy, the matrix elements made of Wilson line operators appearing in the OPE,
describe the non perturbative part of the process, and their evolution in rapidity 
is related to the evolution of the structure function of the target.
In order to find the evolution equation one may use the background field technique. 
The Wilson-line operators evolve with respect to rapidity according to the Balitsky equation, which reduces to the BK equation
in the large $N_c$ limit.
The BK equation describes the so called fan diagrams 
neglecting all non-planar contributions, while  the Balitsky equation gives one the possibility to 
describe the fan diagrams including also the non-planar contributions. 

A similar method to the background field technique is given in Ref.~\cite{Shuvaev:2006br}, where 
the authors studied the propagation of a fast moving particle,
and showed that resumming the emission of soft gluons from this source and
neglecting its recoil (eikonal approximation), one may obtain the BFKL equation.

The Color Glass Condensate (CGC) 
\cite{JalilianMarian:1997gr, JalilianMarian:1997dw, JalilianMarian:1997jx, JalilianMarian:1998cb, Kovner:2000pt, Iancu:2000hn, Iancu:2001ad,
Ferreiro:2001qy, Weigert:2000gi} (for a review, see Ref.~\cite{Iancu:2002xk}) is another available way to describe high-energy scattering processes. This approach is similar to the Wilson line formalism described above, and the corresponding evolution equation is
the Jalilian-Marian, Iancu, MacLerran, Weigert, Leonidov, and Kovner (JIMWLK) equation. 
Indeed, the Balitsky equation, as already mentioned above, is not a closed evolution equation, but it contains an infinite set of 
evolution equations which goes by the name of Balitsky-hierarchy. 
The JIMWLK equation is equivalent to this hierarchy of evolution equations thus leading to the acronym B-JIMWLK equation. 

In all of these approaches,
the  description of the scattering of two probes involves  the computation of the interaction of one Wilson loop describing one probe with the field of the other.
Note that within the dipole model, a similar approach, based on 
the computation of the scattering phase of a dipole in the field emmited
by a fast moving object (involving color structures as well as multicolor states), was  suggested in Ref.~\cite{Kovchegov:1997dm}.

A further approach to the description of high-energy processes is based on an effective field theory. 
This formalism \cite{Kirschner:1994gd, Kirschner:1994xi, Lipatov:1995pn, Antonov:2004hh} provides
the building blocks (reggeon-reggeon-gluon)
necessary to the explicit computation of any type of 
diagram at this regime. However, the precise relationship between this effective theory and the EGLLA approach has not been clarified yet, 
and explicit applications of this effective field theory is highly desirable 
\cite{Braun:2006sk, Hentschinski:2008im, Braun:2009ki}.

The triple $\pom$omeron vertex is the first non-trivial  building block common to all the above approaches. 
It turns out that up to now, its exact expression, including planar and non-planar contribution, was derived only in the EGGLA approach \cite{Bartels:1994jj}. Using its conformal invariance \cite{Bartels:1995kf}, 
both the planar and the non-planar contributions of the vertex were computed in the coordinate space \cite{Lotter:1996vk} (see Ref.~\cite{Korchemsky:1997fy} for 
explicit expressions in the $SU(N_c)$ case). 

So far, only the 
planar contribution was derived with formalisms different from the EGLLA one. These are the dipole model 
\cite{Peschanski:1997yx, Bialas:1997ig, Bialas:1997xp} and the  Wilson line formalism applied to diffractive processes \cite{Balitsky:1997mk}. 

The purpose of the present paper is to show that the exact expression of the 3 $\pom$omeron vertex (planar and non-planar contributions)
can be easily derived through the Wilson line formalism not only for diffractive case but also for fan diagrams. 

Indeed, this formalism was already used
to derive several new and desirable results, and to confirm others which have been derived after many years of calculations. This include for example the NLO BK kernel in 
QCD \cite{Balitsky:2008zza} and in $\cal N$=4 SYM \cite{Balitsky:2009xg} whose linearized version confirmed 
the NLO BFKL kernel in QCD \cite{Fadin:1996tb, Camici:1997ij, Fadin:1998py, Ciafaloni:1998gs} and  in $\cal N$=4 SYM \cite{Kotikov:2000pm}. 
Moreover, through Wilson line formalism it was possible to derive for the first time the conformal expression for the NLO BFKL kernel \cite{Balitsky:2008rc}, 
the full amplitude at NLO in $\cal N$=4 SYM for four scalar currents made of chiral-primary operator
\cite{Balitsky:2009yp}, and 
the analytic result for the NLO photon impact factor \cite{Balitsky:2010ze}, relevant for
phenomenology at high-energy, was recently derived.

It is clear that the Wilson line approach to the study of high-energy scattering processes 
opens the way to attack more difficult problems which have not been solved before despite 
the many efforts that have been devoted to them using different techniques. This includes for example
 multiple  $\pom$omeron or $\odd$dderon vertices, as well as sub-leading contributions. 
 
\section{Introduction to the Wilson line formalism}
\label{Sec: Wilson_approach}

In this section we will give a brief introduction to the Wilson line formalism. This will set the notations
which will be used 
for the derivation of
the  triple $\pom$omeron vertex first in the diffraction case and then for fan diagrams which are important for unitarization.

The main tool which we will use is the OPE for high energies  \cite{Balitsky:1995ub} of the $T-$product of two electromagnetic currents in terms of Wilson lines:

\beq
T j_\mu(x) \,j_\nu(y)  = \int d^2 z_1 \, d^2 z_2 \, I^{LO}_{\mu \nu}(x,y;z_{1\perp},z_{2\perp}) \,
 {\rm Tr} \{ \hat{U}(z_{1\perp}) \,  \hat{U}^{\dagger}(z_{2\perp}) \} + \cdots 
\label{jj}
\eq
with the operator $j_\mu(x) = \bar{\Psi}(x) \gamma_\mu \Psi(x)\,.$
This expansion is in terms of a coefficient function to be identified with the photon impact factor
and a matrix element of two Wilson line operators, with
\begin{equation}
\hat{\cal U}(x_\perp,y_\perp)=1-{1\over N_c}
{\rm Tr}\{\hat{U}(x_\perp)\hat{U}^{\dagger}(y_\perp)\} \,,
\label{defUcal}
\end{equation}
where the Wilson line is defined as usual by the operator
\beq
\label{defWilson}
\hat{U}(x_\perp)={\rm P \,exp}\Big\{ig\int_{-\infty}^\infty\!\!  du ~p_1^\mu 
\, A_\mu(p_1 \, u + x_\perp)\Big\}\,. 
\eq
We use here 
the standard Sudakov decomposition $k = \alpha \,  p_1 + \beta \, p_2 + k_\perp\,,$
where $p_1$ and $p_2$ are light-like vectors defined in such a way that in a typical high-energy scattering process, the first projectile (refered as the ``above'' one) flies almost along $p_1$ while the second one  (resp. ``below'') flies almost along $p_2$.

The operator $\hat{\cal U}$ evolves according to the Balitsky equation
\begin{eqnarray}
&&\hspace{-7mm}
{d\over d\eta}~\hat{\cal U}(x_\perp,y_\perp)=
{\alpha_sN_c\over 2\pi^2}\!\int\!d^2z~ {(x-y)^2_\perp\over(x-z)^2_\perp(z-y)^2_\perp}
[\hat{\cal U}(x_\perp,z_\perp)+\hat{\cal U}(y_\perp,z_\perp)-\hat{\cal U}(x_\perp,y_\perp)-\hat{\cal U}(x_\perp,z_\perp)\, \hat{\cal U}(z_\perp,y_\perp)]\,,\,
\label{bk_U}
\end{eqnarray}
where $x_\perp$ $y_\perp$, $z_\perp$ are two-dimensional vector  with Euclidean metric\footnote{Hereafter, these variables will be denoted simply like $x,$ $y$, $z$.}.
The $\eta$ dependence of the operator $\cal U$ enters as a regulator of the
divergence by  changing the slope of the Wilson line according to
$p_1  \rightarrow  p_1 + e^{-2\eta} p_2$ in Eq.~(\ref{defWilson}).
%

The BK equation \cite{Balitsky:1995ub, Kovchegov:1999yj} is obtained from Eq. (\ref{bk_U}) at large $N_c$ when the correlation function of the non-linear term 
$ \hat{\cal U}(x,z)\, \hat{\cal U}(z,y) $ decouple in a product of two correlation functions 
$\langle \hat{\cal U}(x,z) \rangle \, \langle \hat{\cal U}(z,y) \rangle\,.$ From now on we will
use the short-hand notation 
$
{\cal U}(x, y) \equiv {\cal U}_{xy}\,.
$

\section{Triple $\pom$omeron vertex from diffraction}
\label{Sec:3Pom_diffraction}

\subsection{Diffraction within Keldysh formalism}
\label{SubSec:Keldysh}

We will derive in this section the triple $\pom$omeron vertex for diffractive processes, using Keldysh formalism adapted to describe 
diffractive processes
through functional integration \cite{Balitsky:1988fi}.
The idea is to use  the OPE for diffractive high-energy processes  \cite{Balitsky:1997mk}  in order to reproduce automatically
 the Cutkosky rules for the calculation of total cross-sections.
 One introduces two different fields, each of them living on one side of the cut, which results in three different propagators: $\langle A^+ A^+ \rangle\,,$
$\langle A^+ A^- \rangle\,,$ $\langle A^- A^- \rangle$.

Following \cite{Balitsky:1997mk, Balitsky:2001gj}, we write 
the diffractive amplitude of the $\gamma^* p \to p' + X$ process as 
\begin{eqnarray}
W^{\rm diff}=
\sum_{{\rm flavors}} \! e_{i}^{2} \int {d^{2}k_{\perp}\over 4\pi^2}
   I^A(k_{\perp},0) \,
\langle N|{\rm Tr}\{ \hat{W}^{\zeta=m^2/s}(k_{\perp})
\hat{W}^{\dagger,\zeta=m^2/s}(-k_{\perp})\} |N\rangle,
\label{diffractive_amplitude}
\end{eqnarray}
where $\hat{W}(k_{\perp})$ is the Fourier transform of
\begin{equation}
\hat{W}(x_{\perp})=\hat{V}^{\dagger}(x_{\perp})\hat{U}(x_{\perp}),~~~\hat{W}^{\dagger}(x_{\perp})=
\hat{U}^{\dagger}(x_{\perp})\hat{V}(x_{\perp}) \,,
\label{defW}
\end{equation}
and $\hat{U}(x_{\perp})$ now denotes the  Wilson-line operator 
constructed from $A^+$ fields while $\hat{V}(x_{\perp})$ denotes the 
same operator 
constructed from $A^-$ fields:
\begin{eqnarray}
\hat{U}(x_\perp)&=&{\rm P \, exp}\Big\{ig\int_{-\infty}^\infty\!\!  du ~p_1^\mu 
~A^+_\mu(p_1 \, u + x_\perp)\Big\}\nonumber\\
\nonumber\\
\hat{V}(x_\perp)&=&{\rm P\, exp}\Big\{ig\int_{-\infty}^\infty\!\!  du ~p_1^\mu 
~A^-_\mu(p_1 \, u + x_\perp)\Big\}\,.
\label{defU_V}
\end{eqnarray}
In Eq.~(\ref{diffractive_amplitude}) $m$ is the mass of the proton, and the notation $\zeta = e^{-2\eta}$ is used.
In the case of diffractive processes, the operator 
\begin{equation}
\hat{\cal W}(x_\perp,y_\perp)=1-{1\over N_c}
{\rm Tr}\{\hat{W}(x_\perp)\hat{W}^{\dagger}(y_\perp)\}
\end{equation}
evolves according to the same Balitsky equation (\ref{bk_U}), as \cite{Balitsky:1997mk}
\begin{eqnarray}
&&\hspace{-7mm}\frac{d}{d\eta} \hat{{\cal W}}(x_{\perp},y_{\perp})=
\label{dmaster}
{\alpha_s N_c \over 2\pi^2}\int d^2z
{(x-y)^2_\perp\over(x-z)^2_\perp(z-y)^2_\perp}\Big\{\hat{{\cal W}}(x,z)
+\hat{{\cal W}}(z,y)-\hat{{\cal W}}(x,y)-
\hat{{\cal W}}(x,z)\hat{{\cal W}}(z,y)\Big\}
\,.
\label{eqn_calW}
\end{eqnarray}
In the RHS of this equation, the non-linear
term should be interpreted as the splitting of a diffractive
$\pom$omeron defined by $\langle  \hat{{\cal W}}(x_{\perp},y_{\perp}) \rangle\,.$
We now define ${\cal \hat{V}}$ in a similar way as ${\cal \hat{U}}$ in Eq.(\ref{defUcal})
\begin{equation}
\hat{\cal V}(x_\perp,y_\perp)=1-{1\over N_c}
{\rm Tr}\{\hat{V}(x_\perp)\hat{V}^{\dagger}(y_\perp)\} \,.
\label{defVcal}
\end{equation}
Our goal is now to extract from the non-linear part in Eq. (\ref {dmaster}) terms of the type $\langle {\cal \hat{U}} \rangle \,  
\langle {\cal \hat{V}} \rangle$ which will be interpreted respectively as the $\pom$omeron
on the left (resp. right) of the cut. In order to do this, we should linearize this non-linear term  up to two gluons accuracy\footnote{In the following, the notation $\hat{}$ on operators will be removed for simplification.}.


\subsection{Linearization of non-linear term}
\label{SubSec:linearization_non-linear}

The idea of linearization consists in expanding the non
 linear term of Eq. (\ref{eqn_calW})
up to $g^4$ and rewriting this result in terms of products of the type 
${\cal U} \,  {\cal V}\,.$
The two gluon approximation  means that each ${\cal U}$ and ${\cal V}$
should be approximated up to $g^2\,,$ since each $\pom$omeron
is a singlet color object. The obtained result, whose detailled derivation is given in the Appendix,  is:
\begin{eqnarray}
\label{linnonlinterm}
\hspace{-.4cm}&&{\rm Tr}\{W_xW_z^\dagger-1\}{\rm Tr}\{W_zW_y^\dagger-1\}\stackrel{2g}{=}
\calu_{xz}\calv_{zy} + \calu_{yz}\calv_{zx} +
{1\over N_c^2-1}\left[\calu_{xy} -\calu_{xz}-\calu_{yz}\right]\left[\calv_{xy} -\calv_{zx}-\calv_{zy}\right]\nonumber\\
&&\hspace{-.45cm}
=\!{N_c^2\over N_c^2-1}\!\Bigg[\calu_{xz}\calv_{zy} + \calu_{yz}\calv_{zx} 
+{1\over N_c^2}\Big[\!-\calu_{xy}\calv_{xy}+\calu_{xz}\calv_{xz}+\calu_{zy}\calv_{zy} +\calu_{xy}\Big(
\calv_{xy}-\calv_{xz}-\calv_{zy}\Big)  \nonumber
\\
&&\hspace{-.45cm}
+ \calv_{xy}\Big(
\calu_{xy}-\calu_{xz}-\calu_{zy}\Big)\Big]\!\Bigg] \!.\!\!
\end{eqnarray}


\subsection{Projection on BFKL Green functions}
\label{SubSec:projection_BFKL}

\begin{figure}[h]
\psfrag{Pc}[cc][cc]{\raisebox{.5cm}{\scalebox{1.5}{$\tilde{\Psi}'_{\rm amp.}$}}}
\psfrag{V}[cc][cc]{\raisebox{.5cm}{\scalebox{1.5}{$V$}}}
\psfrag{Pa}[cc][cc]{\raisebox{.5cm}{\scalebox{1.5}{$\tilde{\Psi}$}}}
\psfrag{Pb}[cc][cc]{\raisebox{.5cm}{\scalebox{1.5}{$\tilde{\Psi}$}}}

\centerline{
\includegraphics[width=5.8cm]{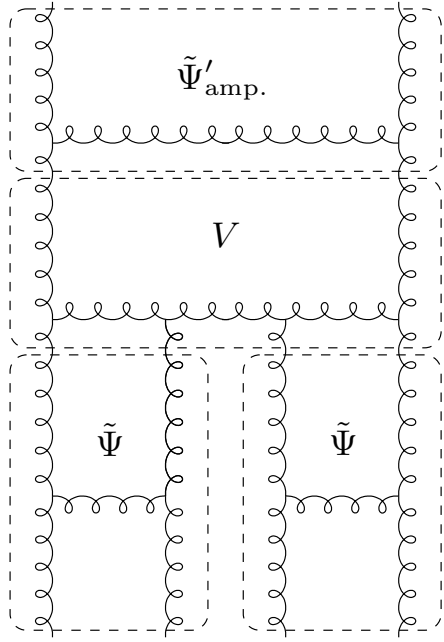}
}
\caption{A typical diagram contributing to the triple $\pom$omeron vertex $V,$ with three 4-gluon Green functions 
(denoted as $\tilde{\Psi}$ and $\tilde{\Psi}'$ in the Moebius representation) at $g^2$ order. 
This exhibits explicitly the amputation of the above Green function from below.}
\label{Fig:3PomFactorized}
\end{figure}

In order to extract the triple $\pom$omeron vertex, one has to define precisely
 how
to factorize out each of the three $\pom$omeron Green function
 from 
the  three-$\pom$omeron correlator. This is achieved according to
 Fig.
\ref{Fig:3PomFactorized}. We denote the above (below)  $\pom$omeron Green functions by $\tilde{\Psi}'$ (resp. $\tilde{\Psi}$).
The vertex $V^{1 \pom \to 2 \pom}$ is defined symbolically in the following way:
\beq
\label{defVop}
\langle \pom \, \pom \, \pom \rangle= (\Delta \Delta \tilde{\Psi}') V^{1 \pom \to 2 \pom} \tilde{\Psi} \tilde{\Psi} 
\eq
where $\Delta \Delta \tilde{\Psi}'$ is the amputated $\pom$omeron Green function which is denoted as 
$\tilde{\Psi}'_{\rm amp.}$ in  Fig.
\ref{Fig:3PomFactorized}.
%
This prescription is in accordance with standard definitions, as exhibited for example in Ref. \cite{Bartels:2004ef}, from which notations of Fig. \ref{Fig:3PomFactorized} are inspired.
We will now translate this definition of the triple $\pom$omeron vertex in the shock-wave approach. 

In the present treatment, we deal with colorless probes. These probes are 
dipoles, which respect the global conformal invariance of the BFKL equation. 
The dipole-dipole scattering, in the BFKL approximation, can then be presented as an elementary function of a conformal anharmonic ratio. This is the basis of the so-called Moebius representation
of BFKL.

In the shock-wave analysis, $\langle \calu_{xy} \rangle$ describes the two-gluon non-amputated amplitude in the Moebius representation (the upper probe is a dipole with coordinate $x$ and $y$ referring respectively to the position of the quark and anti-quark pair),
where the average is on the external field of a lower probe. The contact with the usual 4-gluon BFKL Green function in the Moebius representation can be made if 
one considers the specific case of a lower probe made of two Wilson lines in a color singlet state (a dipole), each of them having definite transverse coordinates.

The dynamics of the process is encoded in the dipole kernel (identical to the BK kernel) which acts on the coordinates $x$ and $y$ 
(the same situation appears for the evolution of multiple dipole densities in the dipole model, 
which is the starting point of the Kovchegov approach), while the averaging from 
below on a given probe does not affects this dynamics. In both the BFKL  and the shock-wave pictures, 
this dynamics is encoded in a kernel which acts on non-amputated functions 
(this remains true also in the dipole picture). 

Now, on one side, the BFKL Green function is non-amputated, 
both from below and from above. Indeed, at Born order it
simply reduces to the product of two propagators. Its Moebius representation is obtained by Fourier transform 
and then by a ``substraction``, which is needed to enforce its vanishing for equal upper or lower coordinates \cite{Bartels:2004ef}. 
This can be obtained directly when computing the elementary dipole-dipole scattering amplitude 
in the two gluon exchange approximation \cite{Navelet:1997tx}). On the other side, in both the shock-wave 
and dipole formalisms, the operator $\calu$ or the dipole densities correspond to amputated quantities 
from the point of view of the below probe. 

Indeed, to get
a scattering amplitude one should convolute these amputated Green function to 
the below probe, thus restoring the gluonic propagators, as it is indeed done in these 
two formalisms. In the Wilson line formalism the fields are not contracted with the probe from below (no propagators). 
In the dipole model, when computing the scattering of two onia, 
in the frame where the below onium is almost at rest, one first evaluates the 
dipole content of the above probe. Then, the restoration of these gluonic propagators appears 
when convoluting this dipole density with the elementary dipole-dipole scattering amplitude
between an internal dipole constituent of the above probe and the below dipole. 

\subsection{Projection on conformal three-point functions}
\label{SubSec:projection_conformal}

\begin{figure}[h]
\includegraphics[width=0.55\textwidth]{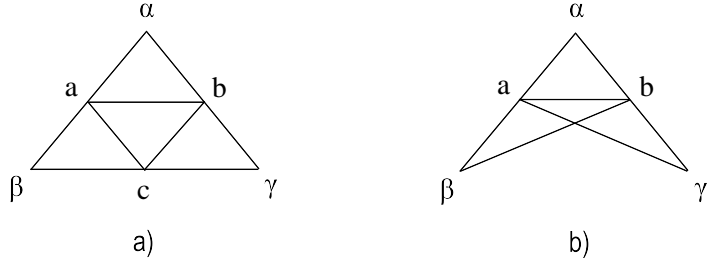}
\caption{a) Configuration of planar diagrams. b) Configuration of non-planar diagrams. \label{Fig:planar-non-planar}}
\end{figure}

In the Moebius representation  the Hamiltonian of the BFKL equation acts as an
integral operator, according to
\begin{eqnarray}
\pi~H^{BFKL} ~f(x,y) = \int d^2 z {(x-y)^2\over (x-z)^2(y-z)^2}\left[
f(x,y)-f(x,z)-f(z,y)\right]\,.
\label{defH}
\end{eqnarray}
Based on the conformal invariance of this BFKL hamiltonian, one can diagonalize it using the following set of eigenfunctions
\begin{eqnarray}
E_{h\bar{h}}(\rho_{10},\rho_{20})\equiv \left(\rho_{12}\over \rho_{10}\,\rho_{20}\right)^h
\left({\bar{\rho}_{12}\over \bar{\rho}_{10}\,\bar{\rho}_{20}}\right)^{\bar{h}}\,,
\label{defE}
\end{eqnarray}
where the conformal weights $h$ and $\bar{h}$ are given by
\beq
h = \frac{1+n}{2} + i \, \nu \,, \qquad \bar{h} = 1 - h^* = \frac{1-n}{2} + i \, \nu \,.
\label{defh}
\eq
The eigenvalue equation then reads
\beq
-{\bar{\alpha}_s\over 2}H^{BFKL}_{xy} E_{h\bar{h}} = 
\omega(h)
E_{h\bar{h}}\,,
\label{diagH}
\eq
with $\bar{\alpha}_s \equiv {\alpha_s N_c\over \pi}\,,$
and
where $\omega(h)$ is given by
\beq
\label{def_omega_h}
\omega(h)= 2 \, \bar{\alpha}_s \, {\rm Re} [\psi(1)-\psi(h)] \,.
\eq
The corresponding BFKL Green function then reads, in the Mellin space $\omega$ conjugated to $s$,
%
\beq
G^{\rm BFKL}_\omega(x, \, y, \, x',\, y') =
\sum_h
\left|\frac{h-\frac{1}{2}}{h(h-1)}\right|^2
\frac{1}{\omega-\omega(h)} \int d^2 \rho_0 \,
E_{h\hb}(x-\rho_{0},y-\rho_{0})\,
E_{h\hb}^*(x'-\rho_{0},y'-\rho_{0}) \,.
\label{Green}
\eq
Let us now turn to the shock-wave formalism. In the BFKL approximation,
the average of the $\calu$ operator on a given probe (denoted as ${\rm B}$) can be written  as
\beq
\label{U_average}
\langle \,{\rm B} | \, \calu_{xy}^{(\omega)}| {\rm B} \rangle= \sum_\alpha \langle x, y \, | \, \alpha \rangle \, \langle \alpha \, | \, {\rm B}  \rangle \,
\frac{1}{\omega-\omega(h_\alpha)}\left|\frac{h_\alpha-\frac{1}{2}}{h_\alpha(h_\alpha-1)}\right|^2\,,
\eq
where the notation $|\alpha \rangle$ is used in order to label both the center of mass $\rho_\alpha$ and the conformal weights $(h_\alpha, \, \bar{h}_\alpha)$. 
Note that in Eq.~(\ref{U_average})
we make an identification of the evaluation of  the operator $\calu$ 
on the state B through functional integration on the LHS, with quantum mecchanical notations appearing on the RHS. Here
$ | x, y \, \rangle$ denotes an upper dipole, $x$ and $y$ being the coordinates of the two corresponding Wilson line. 
The relationship with the 4-reggeon BFKL Green function $G^{\rm BFKL}(x, \, y, \, x',\, y')$  in the Moebius representation is obtained when choosing the lower state B to be a dipole of coordinates $(x',y')\,,$
for which  (\ref{Green}) reads, in the above notations,
\beq
\label{U_average_dipole_down}
\langle \, {\rm dipole}\,(x',y') \, | \, \calu_{xy}^{(\omega)} | \, {\rm dipole}\,(x',y') \, \rangle= \sum_\alpha \,\langle x, y \, | \, \alpha \rangle \, \langle  \alpha \, | \, x', y' \, \rangle \,
\frac{1}{\omega-\omega(h_\alpha)}\left|\frac{h_\alpha-\frac{1}{2}}{h_\alpha(h_\alpha-1)}\right|^2\,,
\eq
leading to the following natural identification:
\beq
\label{E_alpha}
E_{h_\alpha,\hb_\alpha}(x-\rho_\alpha,\,y-\rho_\alpha)=\langle x, y \, | \, \alpha \rangle\,, \qquad 
E_{h_\alpha,\hb_\alpha}^*(x'-\rho_\alpha,\,y'-\rho_\alpha)=\langle  \alpha \, | \, x', y' \, \rangle \,.
\eq

We denote the quantum numbers of the upper $\pom$omeron by $\alpha\,,$ while
the lower left $\pom$omeron (resp. lower right) is labelled by $\beta$ (resp. $\gamma$) (see Fig.~\ref{Fig:planar-non-planar}). 
We are interested here in the extraction of the triple $\pom$omeron vertex.
For that purpose, based on the factorised form (\ref{U_average_dipole_down}) of the BFKL Green function as a series of products of 
conformal blocks, 
the two lower $\pom$omerons are described through $\langle x, z \, | \, \beta \, \rangle$ and $\langle z, y \, | \, \gamma \,\rangle$ while the above one is described through 
$\langle  \alpha \, | \, x , y \rangle\,.$ Now, using the property that 
\beq
\label{E*}
(E_{h,\hb}(x-\rho,y-\rho))^*=E_{1-h,1-\hb}(x-\rho,y-\rho) \,,
\eq
we will not make any distinction between the above and the two below $\pom$omerons,
which will all be described by a set of $E_{h,\hb}$ functions, as it is done for example
in Ref.\cite{Korchemsky:1997fy} to which we would like to compare our final results.


We now fix our notation for the internal degrees of freedom of the vertex, 
based on the non-linear term of Eq.~(\ref{eqn_calW}).
Let us equate the coordinates $x, y$ and $z$ with  $\rho_a, \rho_b$ and $\rho_c$ respectively. It is
convenient to
identify the component of the dipole with definite conformal weight with $\calu\,,$ according to
%
\begin{eqnarray}
&&\calu_{xy} \rightarrow E_{h_\beta\bar{h}_\beta}(\rho_{a\beta},\rho_{b\beta}), ~~~~ 
\calu_{xz}  \rightarrow E_{h_\beta\bar{h}_\beta}(\rho_{a\beta},\rho_{c\beta})\nonumber\\
&&\calv_{xy} \rightarrow E_{h_\gamma\bar{h}_\gamma}(\rho_{a\gamma},\rho_{b\gamma}), ~~~~ 
\calv_{yz} \rightarrow E_{h_\gamma\bar{h}_\gamma}(\rho_{b\gamma},\rho_{c\gamma})\,.
\label{eigenf}
\end{eqnarray}
Now, since we are in the Moebius representation, for each  of the three
$\pom$omeron which are here involved, the Green function is the sum of products of conformal 
block with even conformal spin $n$. This is due to the fact that 
\beq
\label{dipole_BFKL}
G_{\rm Moebius}^{\rm BFKL}(x, \, y, \, x',\, y') \propto G^{\rm BFKL}(x, \, y, \, x',\, y') + G^{\rm BFKL}(y, \, x, \, x',\, y')\,,
\eq
as discussed in Ref.~\cite{Navelet:1997tx} (see Eqs.~(2.50) and (2.51), as well as Eq.~(B.12) in the Born approximation). Therefore, since $n=h+\hb$ is even, and using
the property
\beq
\label{symE}
E_{h,\hb}(x-\rho,y-\rho)=(-1)^{h+\hb}E_{h,\hb}(y-\rho,x-\rho) \,,
\eq
one can make the following identification
\beq
\calu_{ab}= \calu_{ba}= E_{h_\alpha\bar{h}_\alpha}(\rho_{a\alpha},\rho_{b\alpha})
=E_{h_\alpha\bar{h}_\alpha}(\rho_{b\alpha},\rho_{a\alpha})
\label{eigenf_ident}
\eq
where $\calu_{ab}= \calu_{ba}$  in the two-gluon approximation.
Identifications similar to Eq.~(\ref{eigenf_ident}) are done also for $\calu_{bc}$, 
$\calv_{ab}$ etc...

Since the upper $\pom$omeron is amputated, one needs to evaluate the effect of the amputation on a conformal block. This is obtained using the following identity:
\begin{eqnarray}
\Delta_a \Delta_b E_{h_\alpha\bar{h}_\alpha}(\rho_{a\alpha},\rho_{b\alpha})=
16 \, h_\alpha(h_\alpha-1)\bar{h}_\alpha(\bar{h}_\alpha-1)E_{h_\alpha\bar{h}_\alpha}(\rho_{a\alpha},\rho_{b\alpha}){1\over \rho_{ab}^2\,\bar{\rho}_{ab}^2}\,.
\end{eqnarray}

The three-pomeron vertex is now defined, according to Eq.~(\ref{linnonlinterm}), as
\begin{eqnarray}
&&V^{1 \pom \to 2 \pom}=\half\int d^2\rho_a d^2\rho_b d^2\rho_c ~ 16 \, h_\alpha(h_\alpha-1)\bar{h}_\alpha(\bar{h}_\alpha-1)E_{h_\alpha\bar{h}_\alpha}(\rho_{a\alpha},\rho_{b\alpha})
{1\over \rho_{ab}^2\bar{\rho}_{ab}^2}
{|\rho_{ab}|^2\over |\rho_{ac}|^2|\rho_{bc}|^2}
\Bigg[\calu_{ac}\calv_{cb} + \calu_{bc}\calv_{ca} \nonumber\\
&&+{1\over N_c^2}\Big[-\calu_{ab}\calv_{ab}+\calu_{ac}\calv_{ac}+\calu_{cb}\calv_{cb} +\calu_{ab}\Big(
\calv_{ab}-\calv_{ac}-\calv_{cb}\Big) + \calv_{ab}\Big(
\calu_{ab}-\calu_{ac}-\calu_{cb}\Big)\Big]\Bigg]\,.
\label{3p1}
\end{eqnarray}
Using property (\ref{eigenf_ident}), the planar part in Eq.(\ref{3p1}) can be rewritten as
\begin{eqnarray}
\label{Vplanar}
&&
V^{1 \pom \to 2 \pom}_{\rm planar}\\
&&=16 \, h_\alpha(h_\alpha-1)\bar{h}_\alpha(\bar{h}_\alpha-1)\int d^2\rho_a d^2\rho_b d^2\rho_c ~ E_{h_\alpha\bar{h}_\alpha}(\rho_{b\alpha},\rho_{a\alpha})
\frac{1}{|\rho_{ab}|^2 |\rho_{ac}|^2|\rho_{bc}|^2}
E_{h_\beta\bar{h}_\beta}(\rho_{a\beta},\rho_{c\beta})E_{h_\gamma\bar{h}_\gamma}(\rho_{c\gamma},\rho_{b\gamma})\,.\,
\nonumber 
\end{eqnarray}

Let us now consider the non-planar part of the vertex. 
Using the hermiticity property of $H^{\rm BFKL}$ (\ref{defH}) when acting on the product of the two below Green functions $\calu_{xy}\calv_{xy}$ and acting on the above amputated Green function we have,
\begin{eqnarray}
&&\hspace{-2cm}\pi\int d^2\rho_a \,d^2\rho_b \,d^2\rho_c ~
E_{h_\alpha\bar{h}_\alpha}(\rho_{a\alpha},\rho_{b\alpha})~
[ H^{\rm BFKL}_{ab}~\calu_{ab}\calv_{ab}]\nonumber\\
&&\hspace{-2cm}= 
\pi\int d^2\rho_a \,d^2\rho_b \,d^2\rho_c   ~[H^{\rm BFKL}_{ab}
E_{h_\alpha\bar{h}_\alpha}(\rho_{a\alpha},\rho_{b\alpha})]
E_{h_\beta\bar{h}_\beta}(\rho_{a\beta},\rho_{b\beta})E_{h_\gamma\bar{h}_\gamma}(\rho_{a\gamma},\rho_{b\gamma})\nonumber\\
&&\hspace{-2cm}= 
\pi\int d^2\rho_a \,d^2\rho_b  ~ E_{h_\alpha\bar{h}_\alpha}(\rho_{a\alpha},\rho_{b\alpha})~
E_{h_\beta\bar{h}_\beta}(\rho_{a\beta},\rho_{b\beta})E_{h_\gamma\bar{h}_\gamma}(\rho_{a\gamma},\rho_{b\gamma})
\left[-{2\over \bar{\alpha}_s}\omega(h_\alpha)\right]
\end{eqnarray}
Similarly, for the other two terms of the non-planar part we have
\begin{eqnarray}
&&
\int  d^2\rho_c {|\rho_{ab}|^2\over |\rho_{ac}|^2|\rho_{bc}|^2}\left[ \calu_{ab}\Big(\calv_{ab}-\calv_{ac}-\calv_{cb}\Big) + 
\calv_{ab}\Big(\calu_{ab}-\calu_{ac}-\calu_{cb}\Big)\right] 
=\pi ~\calu_{ab} [H_{ab}^{BFKL}\calv_{ab}] + \pi \calv_{ab}[H_{ab}^{BFKL}\calu_{ab}] \nonumber\\
&&=
\pi\Bigg[ E_{h_\beta\bar{h}_\beta}(\rho_{a\beta},\rho_{b\beta}) H^{BFKL}_{ab}~
E_{h_\gamma\bar{h}_\gamma}(\rho_{a\gamma},\rho_{b\gamma})
+E_{h_\gamma\bar{h}_\gamma}(\rho_{a\gamma},\rho_{b\gamma})H^{BFKL}_{ab}~
E_{h_\beta\bar{h}_\beta}(\rho_{a\beta},\rho_{b\beta})\Bigg]\nonumber\\
&& = -\pi{2\over \bar{\alpha}_s}\left[\omega(h_\beta) + \omega(h_\gamma)\right]
E_{h_\beta\bar{h}_\beta}(\rho_{a\beta},\rho_{b\beta})
E_{h_\gamma\bar{h}_\gamma}(\rho_{a\gamma},\rho_{b\gamma})
\end{eqnarray}
Thus, using  once more property (\ref{eigenf_ident}), the non-planar part reads
\begin{eqnarray}
&&
\hspace{-.2cm}V^{1 \pom \to 2 \pom}_{\rm non-planar}=-{2\pi\over N_c^2}\int d^2\rho_a \,d^2\rho_b  ~ 16 \, h_\alpha(h_\alpha-1)\bar{h}_\alpha(\bar{h}_\alpha-1)E_{h_\alpha\bar{h}_\alpha}(\rho_{b\alpha},\rho_{a\alpha}) 
\nonumber \\
&&
\times
{1\over |\rho_{ab}|^4}{\rm Re}\{\psi(1) + \psi(h_\alpha) - \psi(h_\beta) - \psi(h_\gamma)\}
E_{h_\beta\bar{h}_\beta}(\rho_{b\beta},\rho_{a\beta})
E_{h_\gamma\bar{h}_\gamma}(\rho_{b\gamma},\rho_{a\gamma})\,.
\label{Vnon-planar}
\end{eqnarray}
Finally, putting together the non-planar part Eq.(\ref{Vnon-planar}) and the planar one Eq.(\ref{Vplanar}), the triple $\pom$omeron vertex is 
\begin{eqnarray}
&&
\hspace{-.2cm}V^{1 \pom \to 2 \pom}=
 \nonumber\\
&&\hspace{-.5cm}
\int d^2\rho_a \, d^2\rho_b  ~
16 \, h_\alpha(h_\alpha-1)\bar{h}_\alpha(\bar{h}_\alpha-1)
 E_{h_\alpha\bar{h}_\alpha}(\rho_{b\alpha},\rho_{a\alpha}) 
\Bigg[\int d^2\rho_c ~ {1\over |\rho_{ab}|^2|\rho_{ac}|^2|\rho_{bc}|^2}
E_{h_\beta\bar{h}_\beta}(\rho_{a\beta},\rho_{c\beta})E_{h_\gamma\bar{h}_\gamma}(\rho_{c\gamma},\rho_{b\gamma})
 \nonumber\\
&&\hspace{-.5cm}
-{2\pi\over N_c^2}{1\over |\rho_{ab}|^4}{\rm Re}\{\psi(1) + \psi(h_\alpha) - \psi(h_\beta) - \psi(h_\gamma)\}
E_{h_\beta\bar{h}_\beta}(\rho_{b\beta},\rho_{a\beta})
E_{h_\gamma\bar{h}_\gamma}(\rho_{b\gamma},\rho_{a\gamma})\Bigg]\,.
\label{our_result}
\end{eqnarray}
We now compare our result (\ref{our_result}) with the one obtained through the reggeon approach of Ref.~\cite{Bartels:1994jj}, and which is written explicitly
for $SU(N_c)$ and for arbitrary conformal weights in Ref.~\cite{Korchemsky:1997fy}. 
Up to a global normalization factor, related to the convention used to define the triple $\pom$omeron vertex, Eq.(\ref{our_result}) agrees with Eq.~(2.1) of 
 Ref.~\cite{Korchemsky:1997fy}. 

\section{Fan diagram approach}
\label{Sec:Fan}

 The diffractive case, discussed in the previous section, implied that the upper $\pom$omeron
was at $t=0$.
In this section instead, we will show how to obtain the triple pomeron vertex without using the Keldysh formalism. 
This will allow us to obtain the triple $\pom$omeron vertex for fan diagrams
which means that every $\pom$omeron is now at arbitrary $t$. Therefore,
we need to work with the 
   Balitsky equation (\ref{bk_U})
which we rewrite here for convenience
\begin{eqnarray}
&&\hspace{-7mm}
{d\over d\eta}~\hat{\cal U}(x,y)=
{\alpha_sN_c\over 2\pi^2}\!\int\!d^2z~ {(x-y)^2_\perp\over(x-z)^2_\perp(z-y)^2_\perp}
[\,\hat{\cal U}(x,z)+\hat{\cal U}(y,z)-\hat{\cal U}(x,y)-\hat{\cal U}(x,z) \,\hat{\cal U}(z,y)]\,.
\nonumber
\end{eqnarray}
It is easy to see that at large $N_c$ limit the correlation function 
$\langle \, \calu (x_\perp,z_\perp)\, \calu(z_\perp,y_\perp)\rangle$ decouples to the product of two correlation functions
$\langle \, \calu (x_\perp,z_\perp)\rangle \, \langle\, \calu(z_\perp,y_\perp)\rangle$ and this non-linear term is interpreted as the splitting of one
$\pom$omeron into two $\pom$omerons.
In this way the Balitsky equation with the truncation of the hierarchy reduces to
the BK equation; 
its non linear term  coincides exactly with the planar part of the triple $\pom$omeron
vertex \cite{Bartels:2004ef}.
The triple $\pom$omeron vertex takes the following form
\begin{eqnarray}
V_{\rm BK}^{1 \pom \to 2 \pom} \propto \int d^2z {(x-y)^2_\perp\over (x-z)^2_\perp(z-y)^2_\perp }\langle \, \calu_{xz}\rangle\langle\,\calu_{zy}\rangle \,.
\end{eqnarray}
Our aim is now to extract from the non-linear term $\calu_{xz}\,\calu_{zy}$, not only the planar contribution to the triple $\pom$omeron vertex but also the non-planar one. We then show that the result
so obtained coincides with the one we obtained for diffractive processes. We will adopt the following procedure. First, we
consider the correlation function of four Wilson lines
i.e. $\langle\,\calu_{xz}\,\calu_{zy}\,\rangle$, we then apply the two gluon approximation to them and finally we
rewrite the contributions thus obtained in terms of decoupled correlation function of the type 
$\langle\,\calu_{xz}\rangle\langle\,\calu_{zy}\rangle$. This method is technically very similar to the one used in Sec.~\ref{Sec:3Pom_diffraction}, the details of which are given in the Appendix.
We first expand each Wilson line operator. In what follows, we use the shorthand
notation $U_x \approx U^{(0)}_x + U^{(1)}_x + U^{(2)}_x + \dots$, 
with $U^{(0)}$ being the zeroth order term of the expansion, $U^{(1)}_x$ the first order term, and so on. Thus, we have
\begin{eqnarray}
&& N_c^2 \langle\,\calu_{xz}~\calu_{zy}\rangle
\label{expansion}\\
&&\hspace{-.2cm}\approx
\langle \,{\rm Tr}\{1 - (U^{(0)}_x + U^{(1)}_x + U^{(2)}_x)(U^{(0)\dagger}_z + U^{(1)\dagger}_z + U^{(2)\dagger}_z)\} \,
{\rm Tr}\{1 - (U^{(0)}_z + U^{(1)}_z + U^{(2)}_z)(U^{(0)\dagger}_y + U^{(1)\dagger}_y + U^{(2)\dagger}_y)\}\rangle\,.\nonumber
\end{eqnarray}
At this point we want to rewrite eq. (\ref{expansion}) as a sum of decoupled correlation functions like
$\langle\,\calu_{xz}\rangle\langle\,\calu_{zy}\rangle$
such that when we apply the $2-gluon$ approximation to them we get back the Eq. (\ref{expansion}).
Since $U^{(0)}=1,$ the only possible contraction in order to produce terms  of the type $\langle\,\calu_{xz}\rangle\langle\,\calu_{zy}\rangle$
are between $U^{(1)}$ terms or between $U^{(2)}$ terms. Contraction between terms of order higher  than 2 would result in remaining multiplicative terms which are not color singlet (these terms are of the type ${\rm Tr}\, U^{(1)}$
which vanish). Other terms involving contraction of gluon fields at the same coordinate will clearly not contribute:  terms like $\calu_{zz}$ vanish.
Finally, the expansion of each Wilson line is needed only up
to second order.
So, it is then easy to see that
\begin{eqnarray}
&&\hspace{-4mm} \langle\,\calu_{xz}~\calu_{zy}\rangle
\\
&&\hspace{-4mm} \stackrel{2g}{\approx}\!\!
{1\over 2(N^2_c-1)}\!\left\{2 \langle\,\calu_{xz}\rangle\langle\,\calu_{zy}\rangle
\!
 +
 \!
 {1\over N_c^2}\Big[2\langle\,\calu_{xy}\rangle
\big(\langle\,\calu_{xy}\rangle - \langle\,\calu_{xz}\rangle - \langle\,\calu_{yz}\rangle\big) +
 \langle\,\calu_{zy}\rangle\langle\,\calu_{zy}\rangle + \langle\,\calu_{xz}\rangle\langle\,\calu_{xz}\rangle
 - \langle\,\calu_{xy}\rangle\langle\,\calu_{xy}\rangle\Big]\!\right\}
 \!.\nonumber
\label{inclusive-3p}
\end{eqnarray}
We can immediately recognize in Eq.~(\ref{inclusive-3p}) the planar contribution $\langle\calu_{xz}\rangle\langle\calu_{zy}\rangle$ which
coincides with the non-linear term in the BK equation and with the planar part of the diffractive triple
$\pom$omeron vertex obtained in the previous section (cf. Eq.~(\ref{diffractive-2g})).
The terms proportional to $N_c^{-2}$ are instead the non-planar contributions which are suppressed in the
limit of $N_c\to\infty$ and which are, therefore,
obtained only from Balitsky equation and not from BK equation.

We now want to compare the two above approaches: the one based on diffractive processes, Eq. (\ref{3p1}),
and the other one based on the fan diagram approach, Eq. (\ref{inclusive-3p}). The first obtained result is
a particular case of the second one, since it was derived for the splitting $\pom$omeron at $t=0\,.$ Indeed,
the second one can be obtained when identifying $\calu$ with $\calv$
since in the fan diagram case one cannot distinguish between the two 
produced $\pom$omerons.
Let us first consider the planar contribution. In the diffractive approach,
the obtained structure (see Eq.~(\ref{linnonlinterm}))
\beq
\frac{N_c^2}{N_c^2-1} [\,\calu_{ac} \calv_{cb} + \calv_{ac} \, \calu_{cb}]
\label{compa_keldysh}
\eq 
is an operator which should be contracted with an external set of $\pom$omeron
states of quantum numbers denoted by $| \beta \rangle$ and $| \gamma \rangle$
(which as already stated above describes both the conformal weight and the center of mass coordinate of the $\pom$omeron state). At this stage, in the case of the diffractive amplitude, these two 
$\pom$omeron states are distinguishable (one is at the left of the cut while the other one is at the right), and after using the symmetry of the integrand under the replacement $a \leftrightarrow b$, the net result reads symbolically
\beq
2\frac{N_c^2}{N_c^2-1} \, \calu_{ac} | \beta \, \rangle \calv_{cb} | \gamma \rangle\,.
\label{compa_keldysh_result}
\eq 
Now, one can make the identification of the $\calu_{ij} | \beta \, \rangle$
and $\calv_{ij} | \beta \rangle$ states, leading to the final result
\beq
2\frac{N_c^2}{N_c^2-1} \,\calu_{ac} | \beta \rangle \, \calu_{cb} | \gamma \rangle\,.
\label{compa_keldysh_result_final}
\eq
On the other hand, from the fan diagram approach, one has (see first term of Eq.~(\ref{inclusive-3p})), using the same overall normalization as in Eq.~(\ref{compa_keldysh}), 
\beq
\frac{N_c^2}{2(N_c^2-1)} \, 2\, \calu_{ac} \, \calu_{cb} | \beta \, \rangle  | \gamma \rangle = \frac{N_c^2}{N_c^2-1} 2 \, (\calu_{ac} | \beta \,\rangle) \, (   \calu_{cb} | \gamma \rangle)
\label{compa_fan}
\eq 
where the factor of 2 on the RHS of Eq. (\ref{compa_fan}) is
due to the two possible contractions. This shows explicitly that both, the  planar result obtained from the general fan diagram case and the one obtained  from a continuation of the $t=0$ diffractive case in Keldysh formalism, are in agreement.
The proof for the non-planar case follows the same line of thinking.

\section{Conclusion}
\label{Sec:Conclusion}

In this paper we have shown that the triple $\pom$omeron vertex, including the planar and the non-planar contribution, can be obtained very easily
within Wilson line formalism. In Ref.  \cite{Balitsky:1997mk}  this was already done for the case of diffractive processes using Keldysh formalism,
but the result obtained there was only for the planar part of the vertex. In section III, 
we have shown how to compute also the non-planar contribution of the diffractive triple $\pom$omeron 
vertex from Wilson line formalism. To this end we considered the generalization of the Balitsky equation
for diffractive process, and using the linearization procedure and the 2-gluon approximation, we have extracted the desired subleading term in $N_c$
of the vertex.

In section IV we have extended the result of section III to the more generic case of fan diagrams, where the 
$\pom$omeron, which split to two other ones, does not need to be at $t=0$. We then showed that the triple pomeron vertex for fan diagrams 
is the same as the one obtained in the diffractive case. 

Since, as we have shown in the present paper, the Wilson line formalism 
allows one to re-derive very easily results that have been obtained after non trivial and lengthy calculation, we 
plan to use it to study other, still unknown and highly desirable results. For example, it will be interesting to compute
the vertex for $\pom\to 3\pom$ \cite{Ewerz:2003an}, $ \pom \to \odd \, \odd$ \cite{Bartels:2004hb}, $ \odd \to \pom \, \odd$ (so far unknown), or more generally $n\pom \to m\pom$ (inaccessible at the moment through reggeon calculus techniques). These non trivial
building blocks will be relevant in order to identify the unknown underlying effective theory
for high-energy scattering processes. This study is in progress.

\subsection*{Acknowledgements}

We warmly thank I.~Balitsky for many inspiring discussions and comments. We thank
G.~P.~Korchemsky for claryfying us the derivation of formula (2.1) in Ref. \cite{Korchemsky:1997fy}.
We also thank J.~Bartels,  L.~N.~Lipatov, 
S.~Munier, B.~Pire and G.~P.~Vacca for discussions.
This work is partly supported by the ANR-06-JCJC-0084 and by the Polish Grant N202 249235.
G.~A.~C. and L.~S. thank the Institute for Nuclear Theory at the University of Washington 
for its hospitality and the Department of Energy for partial support 
during the completion of this work. G.~A.~C. thanks Lawrence Berkeley National Laboratory for support at
the last stage of this work.
 
\section*{Appendix}
\label{Ap:linearization}

Below  we present some details of the linearization procedure which we used to extract the planar and non-planar part of the triple $\pom$omeron
vertex.

Let us consider the non-linear term in Eq. (\ref{eqn_calW})
\begin{eqnarray}
&&\hspace{-1cm} N_c^2 \,{\rm Tr}\{{{\cal W}}(x,z){{\cal W}}(z,y)\}
= {\rm Tr}\{W_xW_z^\dagger-1\}{\rm Tr}\{W_zW_y^\dagger-1\} = 
{\rm Tr}\{V_zV_x^\dagger U_xU_z^\dagger-1\}{\rm Tr}\{V_yV_z^\dagger 
U_zU_y^\dagger-1\}\,.
\label{nonlinterm}
\end{eqnarray}

Our aim is to extract from this expression the contribution of two
non-interacting $\pom$omerons: one  $\pom$omeron built by $A^+$-fields and the  
other one built by $A^-$-fields.
We now approximate Wilson line operators 
$U$ and $V$ up to linear terms in $A^+$ 
and $A^-$ fields (here we use the short-hand notation $iA^+_x\equiv ig\int_0^1 du \, p_1^\mu A^+_\mu \,(p_1u+x_\perp)$ and similarly for $iA^-_x$)
\begin{eqnarray}
&&N_c^2\,{\rm Tr}\{{{\cal W}}(x,z){{\cal W}}(z,y)\} \nonumber \\
&&
\approx {\rm Tr}\{(1+iA^-_z)(1-iA^-_x)(1+iA^+_x)(1-iA^+_z)-1\}{\rm 
Tr}\,\{(1+iA^-_y)(1-iA^-_z)(1+iA^+_z)(1-iA^+_y)-1\}\,.
\label{linear}
\end{eqnarray}

From each trace of Eq. (\ref{linear}) we keep only terms to second order since we are working in the 2-gluon approximation. Thus, we have
\begin{eqnarray}
N_c^2{\rm Tr}\{{{\cal W}}(x,z){{\cal W}}(z,y)\}
&\approx& 
\;{\rm Tr}\{A^-_zA^-_x + A^+_xA^+_z - A^-_zA^+_x + A^-_zA^+_z + A^-_xA^+_x - A^-_xA^+_z\}
\nonumber \\
&&
\times \,
{\rm Tr}\{A^-_yA^-_z - A^-_yA^+_z + A^-_yA^+_y + A^-_zA^+_z - A^-_zA^+_y - A^+_zA^+_y\}
\,.
\label{quadratic}
\end{eqnarray}
The next step of the linearisation procedure up to 2 gluon accuracy consists in  keeping
in the product of the two traces in (\ref{quadratic})  only terms which 
involve two $A^+$ fields with different coordinates and two $A^-$ fields with 
different coordinates. Thus, we obtain that
\begin{eqnarray}
&& N_c^2\,{\rm Tr}\{{{\cal W}}(x,z){{\cal W}}(z,y)\}
\stackrel{2g}{\approx} 
{\rm Tr}\{A^-_xA^-_z\}{\rm Tr}\{A^+_zA^+_y\} + {\rm Tr}\{A^+_xA^+_z\}{\rm 
tr}\{A^-_yA^-_z\} 
+ {\rm Tr}\{A^-_zA^+_x\}{\rm Tr}\{A^-_yA^+_z\} 
\nonumber \\
&&
- {\rm Tr}\{A^-_zA^+_x\}{\rm Tr}\{A^-_yA^+_y\} + {\rm Tr}\{A^-_zA^+_z\}{\rm 
tr}\{A^-_yA^+_y\}
 - {\rm Tr}\{A^-_xA^+_x\}{\rm Tr}\{A^-_yA^+_z\}
+ {\rm Tr}\{A^-_xA^+_x\}{\rm Tr}\{A^-_yA^+_y\} 
\nonumber\\
&&
+ {\rm Tr}\{A^-_xA^+_x\}{\rm 
tr}\{A^-_zA^+_z\} - {\rm Tr}\{A^-_xA^+_x\}{\rm Tr}\{A^-_zA^+_y\}
- {\rm Tr}\{A^-_xA^+_z\}{\rm Tr}\{A^-_yA^+_y\} + {\rm Tr}\{A^-_xA^+_z\}{\rm 
tr}\{A^-_zA^+_y\}\,.
\label{nextstep}
\end{eqnarray}
We now have to rewrite Eq. (\ref{nextstep}) in terms of product of traces involving only $A^+$ and $A^-$ fields, so  we obtain
\begin{eqnarray}
&&N_c^2\,{\rm Tr}\{{{\cal W}}(x,z){{\cal W}}(z,y)\}
\stackrel{2g}{\approx} 
 {\rm Tr}\{A^-_xA^-_z\}{\rm Tr}\{A^+_zA^+_y\} + {\rm Tr}\{A^+_xA^+_z\}{\rm Tr}\{A^-_yA^-_z\} \nonumber \\
&& + \frac{1}{N_c^2-1} \left[ {\rm Tr}\{A^+_xA^+_y\} - {\rm Tr}\{A^+_xA^+_z\}-{\rm Tr}\{A^+_yA^+_z\}  \right]
\left[  {\rm Tr}\{A^-_xA^-_y\} - {\rm Tr}\{A^-_xA^-_z\} - {\rm Tr}\{A^-_yA^-_z\}  \right]\,.
\label{twosinglets}
\end{eqnarray}
The final step is to write Eq. (\ref{twosinglets}) 
in terms of the original operators $\calu$ and $\calv$ such that when we apply to them the 2-gluon approximation we get back Eq. (\ref{twosinglets}).
So, we have
\begin{eqnarray}
N_c^2\,{\rm Tr}\{{{\cal W}}(x,z){{\cal W}}(z,y)\} \stackrel{2g}{\approx} N_c^2
\left( \calu_{xz}\calv_{zy} + \calu_{yz}\calv_{zx} +
{1\over N_c^2-1}\left[\,\calu_{xy} 
-\calu_{xz}-\calu_{yz}\right]\left[\calv_{xy} 
-\calv_{zx}-\calv_{zy}\right] \right)
\label{diffractive-2g}
\end{eqnarray}
which is the linearization we used in Eq.~(\ref{linnonlinterm}).

\bibliographystyle{prsty}


\end{document}